# Kinetics of segregation formation in the vicinity of edge dislocation in fcc metals

A. V. Nazarov (1 and 2), A. A. Mikheev (3), I. V. Ershova (1) and A. G. Zaluzhnyi (1 and 2)

(1) National Research Nuclear University MEPhI, Moscow, Russia
(2) SSC RF Institute for Theoretical and Experimental Physics (ITEP), Moscow, Russia
(3) Moscow State University of Design and Technology, Russia

## Abstract

We use new equations for the interstitial impurity diffusion fluxes under strain to study impurity atom redistribution in the vicinity of dislocations. Two levels of simulation are applied. The first one is evaluation of coefficients that determine the influence of strain tensor components on interstitial diffusion fluxes in FCC structures. For this purpose we have developed a model into the framework of a molecular static method taking into account an atom environment both near the interstitial site and for the saddle-point configuration. The second level is modeling the interstitial segregation formation based on nonlinear diffusion equations that take strains generated by defects into account. The results show that the distributions of the interstitials near the dislocations have quite complicated characters and the vacancy distribution has a qualitatively different character as compared with the carbon distribution.

# I. Introduction

Elastic fields, generated by defects of the structure, influence the diffusion processes. In addition, as a result, it leads to the alteration of the phase transformation kinetic, segregation formation and changes of the system properties [1-4]. Apparently, the first the elastic interaction between a point defect and an edge dislocation has been treated by Cottrell and Bilby [2] under the assumption that the point defect is a center of dilatation in an isotropic continuum. Koehler [3] has treated an additional term in the determination of the vacancy migration energy in a stress field which arises because the vacancy saddle point configuration has lower symmetry than the lattice, and points out that the migration energy is not isotropic in a stress field. Johnson [4] has treated the interactions for a nonspherical defect near a dislocation using isotropic elasticity theory and has determined the first-order shape-effect elastic interaction. In presented work we study redistribution kinetics of defects in the vicinity of dislocations using our theory of diffusion under stress [5-7] and taking into account the strain generated by dislocations.

In the first part of our paper we give a brief consideration to the main features of the theory of diffusion under strain [5-7] and to present the general equations for the fluxes in interstitial alloys under strain. This approach gives the possibility to use the mentioned equations at low temperatures, in conditions where the strain influence on the diffusion fluxes is manifested in maximal degree. We use evaluation results of coefficients that determine the influence of strain tensor components on interstitial diffusion fluxes in FCC structures and diffusion features [7]. The coefficients are very sensitive to the atomic structure in the nearest vicinity of a defect and still more sensitive to the atomic structure of the saddle-point configuration. Therefore, for these purposes we use the advanced model developed by us earlier [8-10].

For an illustration of our approach possibilities in the next part we examine a formation of interstitial atom segregation near the dislocations by applying the general equations for the fluxes in interstitial alloys under strain in case, when the strain is generated by dislocation. Wherein the strain in the vicinity of dislocations outside the core is described by isotropic elasticity theory and dislocation line is parallel to <100> direction. The last part is concerned with a simulation of the vacancy redistribution in the vicinity of dislocations under similar conditions.

## II. General Equations

We propose an original microscopic approach to examine the effect of elastic strain of a general type on diffusion. The key moment of this approach is explicit consideration of the fact that positions as well as potential energies of atoms are changed due to the displacement field. This is considered both for the equilibrium and saddle-point configurations. Consequently, strain changes the local activation barriers. Knowing the changes of the activation barriers it is possible to calculate the jump rates. The rates of atom jumps in different directions define the flux density of the defects, explicit form of which can be derived by calculating the balance of different jumps and their contribution to the flux. We obtain the expressions for the vacancy diffusion fluxes and interstitial diffusion fluxes in FCC and BCC structures [6,7].

The carbon atom jumps give the contribution to the flux in the direction of axis X in fcc metal on condition that atoms have the following initial and final positions: $(0; 0; 0) \rightarrow (\pm a/2; \pm a/2; 0)$ and $(0; 0; 0) \rightarrow (\pm a/2; 0; \pm a/2)$, where $a$ is a lattice parameter (figure 1).

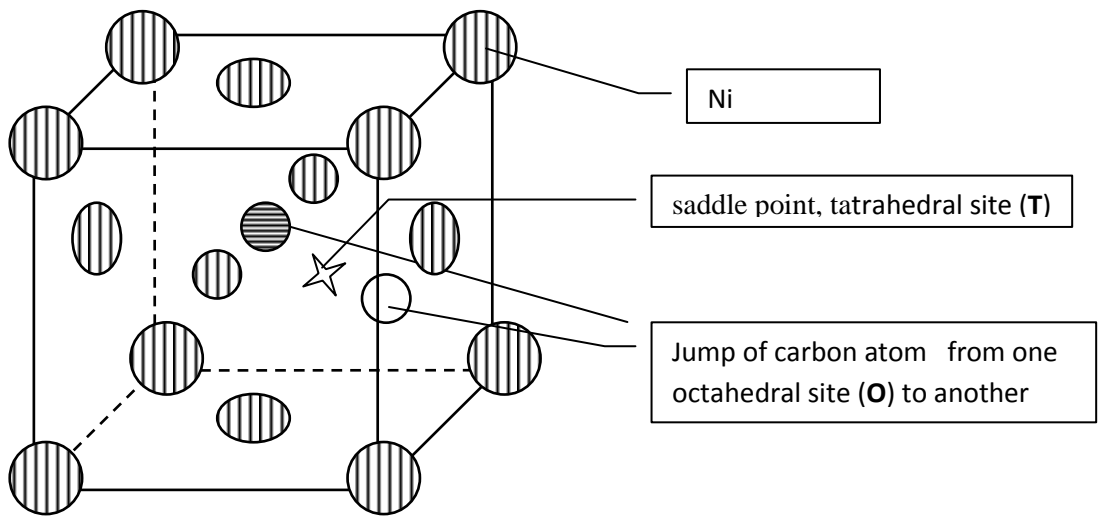

Figure 1. Jumps of the interstitial impurity atom in FCC structures (O – T – O jumps).

For the diffusion of the interstitial atoms in fcc metal under strain twelve (O –T – O) jumps give the contribution to the flux. After calculation of the different jump balance general equations for the X-component of the interstitial atom flux may be written as [7]:

$$J_1 = -\frac{1}{\Omega}\left[\left(D_{11}\frac{\partial c_i}{\partial x}+D_{12}\frac{\partial c_i}{\partial y}+D_{13}\frac{\partial c_i}{\partial z}\right)+c_i\left(\frac{\partial D_{11}}{\partial x}+\frac{\partial D_{12}}{\partial y}+\frac{\partial D_{13}}{\partial z}\right)\right], \qquad (1)$$

where $c_i$ is the interstitial atom concentration, $\Omega$ is the atomic volume.

For the other axes the equations are similar. In this case the coefficients $D_{ij}$ depend on the non-diagonal components of the strain:

$$D_{11}=\frac{D}{2}\exp\left(-\frac{K_1\operatorname{Sp}\varepsilon}{kT}\right)\left[\exp\left(-\frac{K_2\varepsilon_{33}}{kT}\right)\operatorname{ch}\left(\frac{K_3\varepsilon_{12}}{kT}\right)+\exp\left(\frac{K_2\varepsilon_{22}}{kT}\right)\operatorname{ch}\left(-\frac{K_3\varepsilon_{13}}{kT}\right)\right],$$

$$D_{12}=\frac{D}{2}\exp\left(-\frac{K_1 Sp\varepsilon}{kT}\right)\exp\left(-\frac{K_2\varepsilon_{33}}{kT}\right)\operatorname{sh}\left(\frac{K_3\varepsilon_{12}}{kT}\right), \qquad (2)$$

$$D_{13}=\frac{D}{2}\exp\left(-\frac{K_1 Sp\varepsilon}{kT}\right)\exp\left(-\frac{K_2\varepsilon_{22}}{kT}\right)\operatorname{sh}\left(\frac{K_3\varepsilon_{13}}{kT}\right),$$

where $\varepsilon_{ij}$ is the strain tensor, $D$ is the diffusion coefficient of interstitial atoms for the perfect crystal. One can readily see that each of these coefficients depends on strain tensor components in a nonlinear way. In corresponding nonlinear equations the functional dependence of strain is determined by coefficients, which are the main characteristics of the strain influence on diffusion (SID coefficients). In the case of the interstitial atom diffusion in fcc metals for (O – T – O jumps) three coefficients $K_1, K_2, K_3$ determine the influence of strain on diffusion, each of that is a linear combination of the coefficients:

$$K_{xx}^W=\frac{1}{2}\sum_s\sum_{k\neq s}\frac{(x_{ks}^W)^2}{R_{ks}^W}\left.\frac{\partial E}{\partial R_{ks}}\right|_{R_{ks}^W}, \quad K_{xx}^0=\frac{1}{2}\sum_s\sum_{k\neq s}\frac{(x_{ks}^0)^2}{R_{ks}^0}\left.\frac{\partial E}{\partial R_{ks}}\right|_{R_{ks}^0},$$

$$K_{xy}^W=\frac{1}{2}\sum_s\sum_{k\neq s}\frac{x_{ks}^W y_{ks}^W}{R_{ks}^W}\left.\frac{\partial E}{\partial R_{ks}}\right|_{R_{ks}^W} \qquad \text{and analogical them,} \qquad (3)$$

where $x_k, y_k, z_k$ are the coordinates of the atom $k$, $x_{ks}=x_k-x_s$, $y_{ks}=y_k-y_s$, $z_{ks}=z_k-z_s$, $k\neq s$, $R_{ks}=|\mathbf{r}_k-\mathbf{r}_s|=\sqrt{x_{ks}^2+y_{ks}^2+z_{ks}^2}$ for all atoms $k$, $E=E(R_{ks})$ is the system energy without strain. These coefficients depend on the atomic interactions and atomic configurations of the defect environments such as: when the atom comes to the saddle point (relative distances between the atoms have index W) and when the atom is at equilibrium position (relative distances between the atoms have index 0). The effect of elastic stress field on the diffusion coefficient matrix is defined by three coefficients $K_1, K_2, K_3$ for fcc metals:

$$K_1 = K_{yy}^W - K_{yy}^0, \quad K_2 = K_{xx}^W - K_{yy}^W, \quad K_3 = 2K_{xy}^W. \tag{4}$$

Then we use the solution of the classical theory of elasticity for the strain in the vicinity of dislocations outside the core:

$$\varepsilon_{11} = -\frac{b}{4\pi(1+v)} \frac{y}{(x^2+y^2)^2} \left[ \frac{(3+v)}{(1-v)} x^2 + y^2 \right]$$

$$\varepsilon_{22} = \frac{b}{4\pi(1+v)} \frac{y}{(x^2+y^2)^2} \left[ \frac{(1+3v)}{(1-v)} x^2 - y^2 \right]$$

$$\varepsilon_{12} = \frac{b}{4\pi(1-v)} \frac{x(x^2-y^2)}{(x^2+y^2)^2}, \tag{5}$$

where *b* is the Burgers vector of the dislocation and *v* is the Poisson ratio.

Further we substitute the components of the strain tensor equation (5) in equations for $D_{ij}$ (equations (2)). Finally, if we substitute the obtained expressions for fluxes in the continuity equation then we get the diffusion equation, in which the influence of elastic strain on flux is taken into account. In particular, this equation gives the possibility to describe interstitial redistribution under strain.

### III. Models and Results

We model redistribution of impurity atoms with the help of a numerical solution of non-linear diffusion equation taking into account the strain generated by dislocations. In this equation the functional dependence on strain is determined by coefficients, which are the main characteristics of the strain influence on diffusion (SID coefficients). These coefficients are determined with the help of atomic simulation for some interstitial alloys [7] in a similar way like it was done in [8-10] for vacancies. We used N-body potentials developed in [11,12], describing carbon atom interaction with the atom of metals, and our advanced model [8-10], to evaluate these coefficients.

The knowledge of elastic fields let us model the segregation formation kinetics of carbon impurity atoms in the vicinity of dislocations at different temperatures. The uniform initial concentration of impurities was taken and it equals $c_i^0$. The distributions of carbon atom concentration near the dislocation are presented on figure 2.

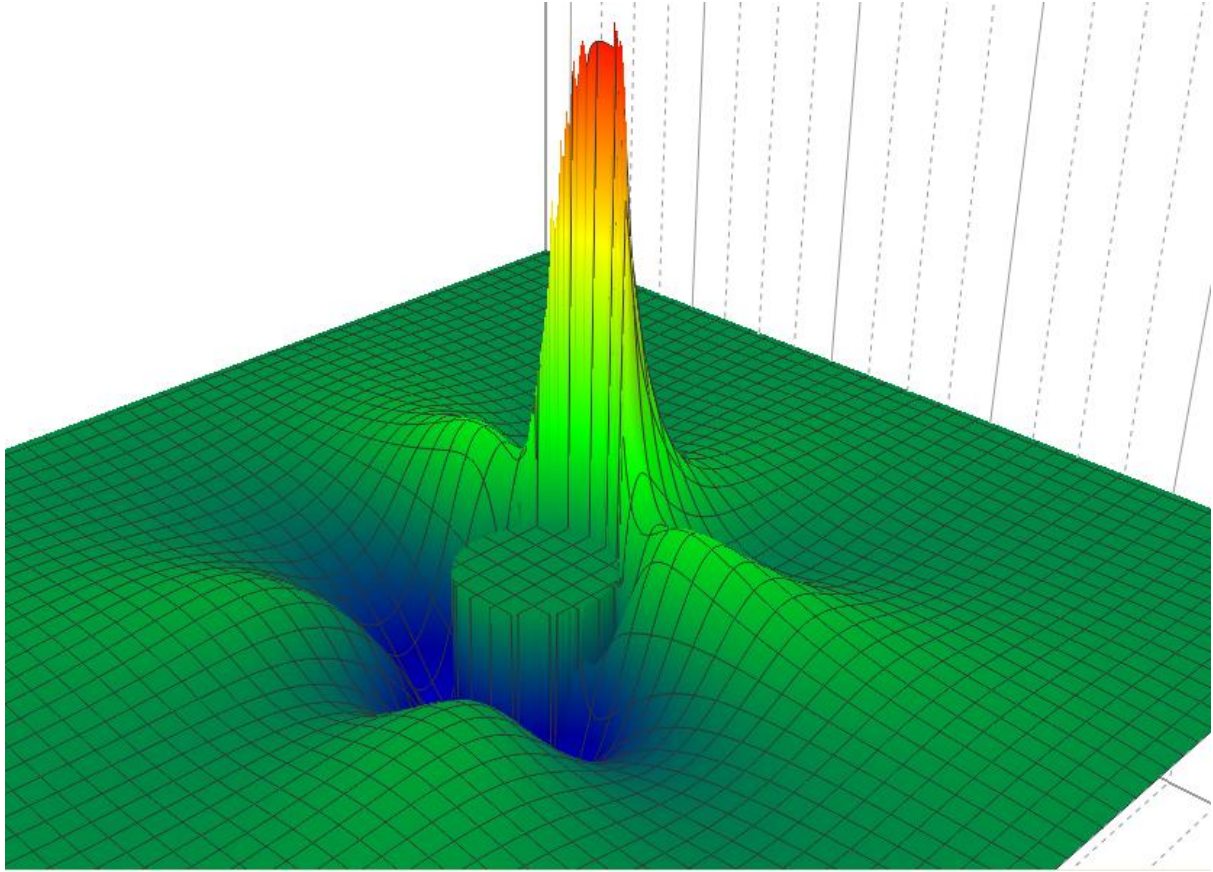

Figure 2. Distribution of carbon impurity atoms in the vicinity of dislocation.
$T= 0.4\ T_\mathrm{m}$, $t = 5\mathrm{d}t$, $c_i^{max}/c_i^0 = 2.1$

Here $T$ is the temperature, $T_\mathrm{m}$ is the melting temperature of metal, $\mathrm{d}t$ is the conventional time unit. The results of this simulation show that the distribution of the interstitials near the dislocations has a quite complicated character. A corresponding function $c_i(x,y)$ in the normal plane to the line dislocations has some ridges and valleys between them. The character of the distribution depends on the elastic field generated by the dislocations, on the SID coefficients and their relation to the activation barrier value for a perfect crystal and on the temperature.

Simulation of the vacancy redistribution in the vicinity of dislocations is realized a similar way. But in this case we used the general equations for the vacancy fluxes in metals under strain [6,7] and SID coefficients calculated for Ni. The results are presented on figures 3 and 4.

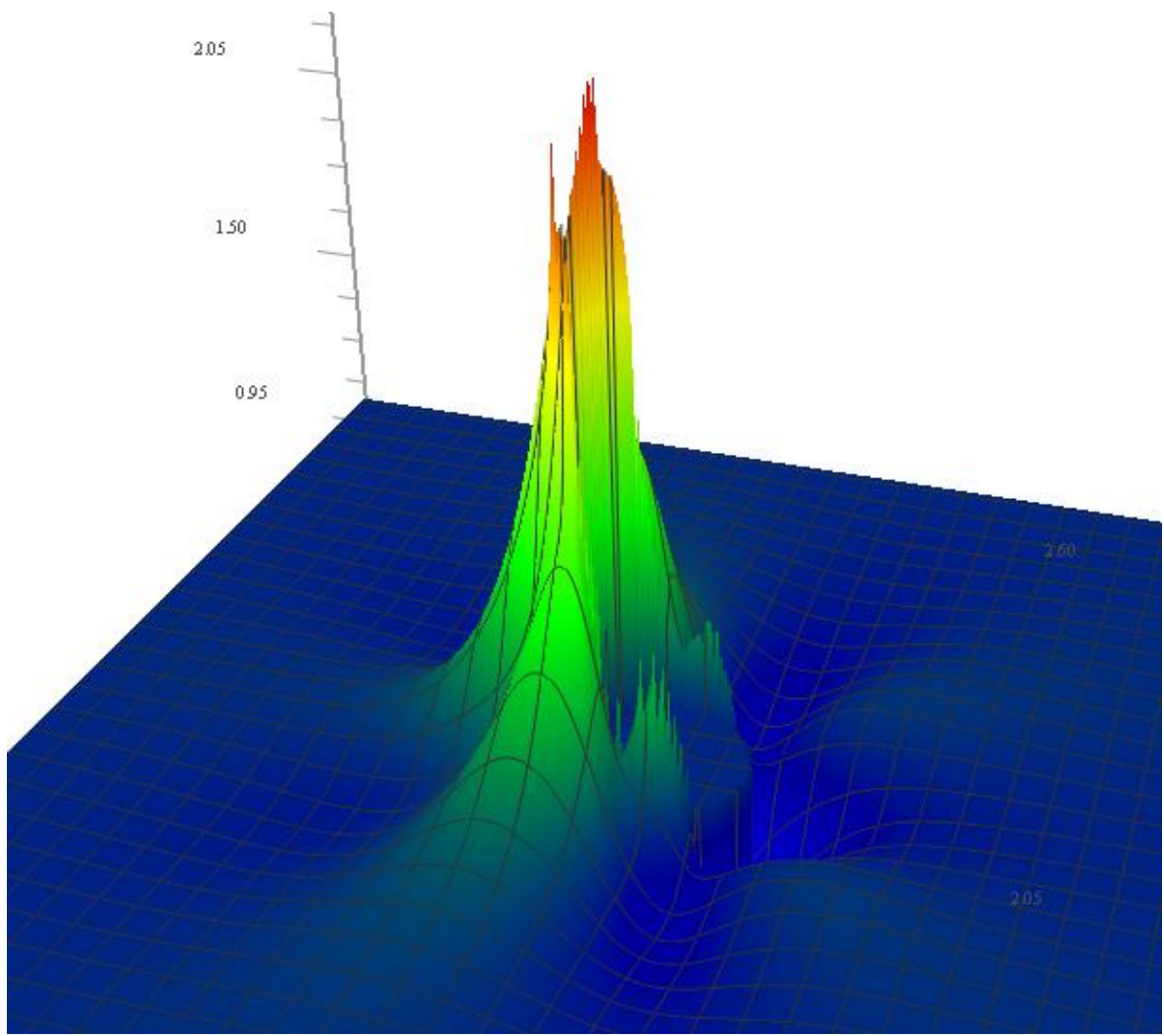

Figure 3. Vacancy distribution in the vicinity of dislocation.
$T = 0.4\ T_\mathrm{m}$, $t = 5\mathrm{d}t$, $c_v^{max}/c_v^0 = 2.58$

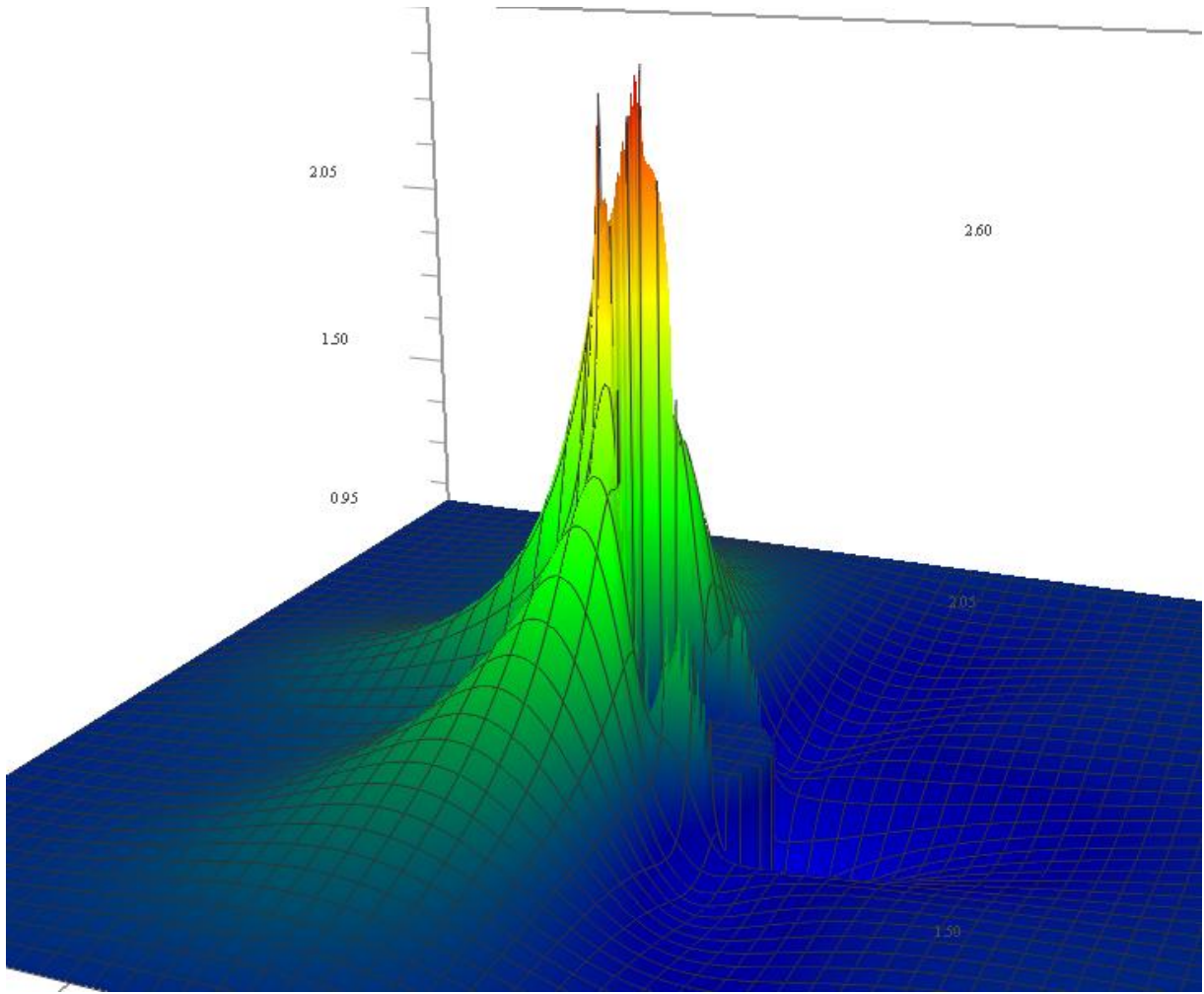

Figure 4. Vacancy distribution in the vicinity of dislocation.
$T= 0.4\ T_m$, $\quad t= 6dt$, $\quad c_v^{max}/c_v^0 = 2.6$

The results of this simulation show that the vacancy distribution near the dislocations differs qualitatively from the interstitial one. It should be noted that in all calculations the concentration near the dislocation core takes a value very close to the maximum in a very short time, and then this value is increased very slowly. The concentration increase is much slower at large distances from the core. This effect is caused by strongly nonlinear equations and can play an important role in the formation of nucleation on dislocations such phases as carbides and hydrides and in the formation of void nucleation on dislocations.

## IV. Conclusion

Based on the theoretical approach, developed earlier, the nonlinear equation for interstitial diffusion under the strain generated by dislocations in fcc metals has been derived.

Multi-scale model of diffusion, which allows to take into account the influence of the strain on kinetics of interstitial atom redistribution near the dislocation is developed.

Simulations of redistribution kinetics near the dislocation are made for carbon impurity atoms in Ni.

Simulations of vacancy redistribution kinetics near the dislocation are made for different temperatures.

The results of the simulation show that the vacancy distribution near the dislocations differs appreciably from the interstitial one and the distributions have the quite complicated character.

## V. References

[1] Was G S 2007 *Fundamentals of Radiation Materials Science. Metals and Alloys* (New York, Springer)

[2] Cottrell A H and Bilby B A, 1949 *Proc. Phys. Soc.* **A 62** 49.

[3] Koehler J, 1969 *Phys. Rev.* **181**, 1015

[4] Johnson R A 1979 *Journal of Applied Physics* **50** 1263

[5] Nazarov A V and Mikheev A A 2004 *Physica Scripta* **T108** 90

[6] Nazarov A V and Mikheev A A 2008 *J. Phys.: Condens. Matter* **20** 485203.1

[7] Nazarov Andrei, Mikheev Alexander, Valikova Irina and Zaluzhnyi Alexander 2011 *Solid State Phenomena* **172-174** 1156

[8] Valikova I V and Nazarov A V 2008 *Def. Diff. Forum* **277** 125

[9] Valikova I V and Nazarov A V 2008 *Phys. of Metals and Metallography* **105** 544

[10] Valikova I V and Nazarov A V 2010 *Phys. of Metals and Metallography* **109** 220

[11] Ruda M, Farkas D and Abriata J 1996 *Phys. Rev.* B **54** 9765

[12] Ruda M, Farkas D and Abriata J 2002 *Scripta Materialia* vol **46** p 349